\def\sles{\lower2pt\hbox{$\buildrel {\scriptstyle <} 
   \over {\scriptstyle\sim}$}}
\def\sgreat{\lower2pt\hbox{$\buildrel {\scriptstyle >} 
   \over {\scriptstyle\sim}$}}
\documentstyle[aps,aipbook,floats,epsf]{revtex}
\begin{document}
\title{Are There MeV Gamma-Ray Bursts?}

\author{Tsvi Piran$^{1,3}$ and Ramesh Narayan$^{2,3}$} 

\address{1. The Racah Institute for Physics, The
Hebrew University, Jerusalem, Israel\\
2. Harvard-Smithsonian Center for
Astrophysics, Cambridge, MA , U.S.A.\\
3. ITP, UCSB,  Santa Barbara, , U.S.A.}

\maketitle
\begin{abstract}
It is often stated that gamma-ray bursts (GRBs) have typical energies
of several hundred keV, where the typical energy may be characterized
by the hardness $H$, the photon energy corresponding to the peak of
$\nu F_\nu$. Among the 54 BATSE bursts analyzed by Band et. al.
more than half have $100{\rm keV} < H < 400$keV. Is the narrow range
of $H$ a real feature of GRBs or is it due to an observational bias?
We consider the possibility that bursts of a given bolometric
luminosity occur with a distribution: $p(H)d \log H
\propto H^\gamma d \log H$. We model the detection
efficiency of BATSE as a function of $H$ and calculate the expected
distribution of $H$ in the observed sample for various values of
$\gamma$.  The Band sample shows a paucity of soft (X-ray)
bursts, which may be real.  However, because the detection efficiency
of BATSE falls steeply with increasing $H$, the paucity of hard bursts
need not be real.  We find that the observed sample is consistent with
a distribution above $H = 100$ keV with $\gamma
\approx 0$ (constant numbers of GRBs per decade of hardness) or even
$\gamma =0.5$ (increasing numbers with increasing hardness). Thus, we
suggest that a large population of unobserved hard gamma-ray bursts
may exist. It is important to extend the present analysis to a larger
sample of BATSE bursts and to include the OSSE and COMPTEL limits. If
the full sample is consistent with $\gamma\ \sgreat\ 0$, then it would
be interesting to look for MeV bursts in the future.
\end{abstract}

One striking feature that is common to all gamma-ray bursts (GRBs) is
the fact that most of the observed photons correspond to low energy
gamma-rays, with energies of a few tens to few hundreds of keV.  While
other features of the bursts, in particular the temporal structure,
vary significantly from one burst to another, this feature seems to be
quite invariant.  One wonders, therefore, whether this is a clue to
the nature of GRBs --- a phenomenon that theorists should strive to
explain --- or if it is just the consequence of an observational bias
against detection of harder or softer bursts.

Band et al. \cite{Band} have analyzed 54 strong GRBs (hereafter the Band
sample), fitting the spectra using a four parameter function:
$$
N(E) = \cases { \big({E \over
100{\rm keV}})^{\alpha} \exp (-{E\over E_0}) & for  
$(\alpha-\beta) E< E_0$ ;\cr 
\big[{(\alpha-\beta) E_0 \over 100{\rm keV}}
\big]^{(\alpha-\beta)} \exp (\beta-\alpha)
\big({E\over 100{\rm keV}}\big)^\beta, & for $E > (\alpha-\beta)E_0.$ 
\cr} 
$$
This function, which provides a good fit to most of the observed
spectra, is characterized by two power laws joined smoothly at a break
energy $H \equiv (\alpha-\beta)E_0$.  For most observed values of
$\alpha$ and $\beta$, $\nu F_\nu \propto E^2 N(E)$ rises below the
break and decreases above it. The break energy $H$ is thus the
``typical'' energy of the observed radiation, in the sense that this
is where the source emits the bulk of its luminosity. $H$ is
correlated with, but not equal to, the hardness ratio which is commonly
used in analyzing BATSE GRBs, namely the ratio of photons observed in
channel 3 to those observed in channel 2. 

Figure 1 shows the distribution of observed values of $H$ in the Band
sample.  There is a clear and marked maximum in the distribution for
$H\sim 200$keV, and most of the bursts lie over the range $100\,{\rm
keV}<H<400\,{\rm keV}$.  Should we, therefore,
conclude that most GRBs have hardnesses around 100--400 keV?

To answer this question we have used a simple model of the BATSE
detector to calculate the {\it expected} hardness distribution of GRBs
detected by BATSE for various assumed {\it intrinsic} hardness
distributions.  We have then compared the expected and observed
distributions to see which intrinsic distributions are consistent with
the data and which are not.  

We assume that the {\it intrinsic} hardness distribution is as follows:
\begin{itemize}\item
All GRBs are adequately described by a Band 
spectrum, characterized by the parameters, $\alpha$, $\beta$, and $H$.
\item
The number of GRBs in a logarithmic interval, 
$d\log H$, varies as: $p(H) d\log H = p_0 H^\gamma d \log H$. 
\item
We also allow for the possibility
that the intrinsic luminosity, $L$,  is correlated with $H$: $
L = L_0 (H/H_0)^\xi$.  
\end{itemize}

We now calculate, for given $\gamma$ and $\xi$, the distribution of
$H$ values of bursts seen by a detector like BATSE.  We make use of
the fact that BATSE triggers on counts in the second and third energy
channels, with photon energies between 50keV and 300keV.  We
calculate, therefore, how many bursts with a given $H$ yield a count
rate in the 50keV to 300keV band that is larger than a threshold,
$C_{min}$.  (We ignore the fact that BATSE's counts in these channels
may correspond to higher energy photons and that the intrinsic
spectrum is related to the observed counts via the DRM, \cite{DRM}).  The
count rate depends on $\alpha$ and $\beta$, in addition to the above
mentioned dependence on $\xi$ and $\gamma$. We use the average values
given by Band et al.  \cite{Band}: $\bar\alpha = -0.73$ and $\bar \beta =
-2.22$. For reasons that will become clear shortly we also use $\tilde
\alpha = -0.41$, which is the average value of $\alpha$ for GRBs with
$50{\rm keV} < H < 300$keV.

The count rate in the 50-300keV energy band from a burst at a distance
$D$ is:
$$
C = { L(H) A \over 4 \pi D^2 } {\cal
C}(50,300,\alpha,\beta,H) = \bigg[ { (L_0/H_0^\xi)  A \over 4 \pi
D^2 } \bigg] {\cal C}(50,300,\alpha,\beta,H) H^\xi \ \ ,
$$
where $A$ is the area of the detector.  The quantity ${\cal C}$
represents the number of counts for every erg of energy incident on
the detector (counts/ergs), and is easily calculated once the shape of
the spectrum ($\alpha$, $\beta$, $H$) and the detector limits (50, 300
keV) are given.  The second equality follows from the assumed
correlation between $L$ and $H$. For simplicity we use here and in the
following a Newtonian geometry and ignore cosmological effects. This
is justified since the Band sample is composed mostly of strong bursts
for which cosmological redshift and spatial curvature effects are
small.  A detailed discussion that includes these effects will be
published elsewhere (Cohen, Narayan \& Piran, 1996).

The  number of bursts with hardness $H$ 
detected by a detector with a limiting sensitivity $C_{min}$ is:
$$
N(H) = \bigg[ {4 \pi \over 3}
\bigg( { (L_0/H_0^\xi)  A \over 4 \pi C_{min} } \bigg)^{3/2} 
p_0 \bigg]
{\cal C}^{3/2}(50,300,\alpha,\beta,H) H^{3\xi/2+\gamma}\ .
$$

Figure 1 depicts $N(H)$, the expected number of detected bursts in a
logarithmic interval of $H$, as a function of $\log(H)$.  We have
shown the calculated curves for $\alpha=\bar
\alpha$ and $\alpha=\tilde \alpha$ and for $\gamma+3\xi/2=0$ and
$\gamma+3\xi/2=0.5$; in all cases, we set $\beta=\bar\beta$.  The
observed distribution corresponding to the Band sample, with
statistical error bars, is also shown on the same plot.

A comparison of the four calculated (or expected) curves with the
observed distribution reveals immediately a paucity of soft bursts.
In all four cases, the number of soft bursts we would have expected to
see is significantly larger than the number actually observed.
Therefore, unless BATSE has an unexpectedly large selection bias
against detecting soft photons (i.e. significantly poorer sensitivity
than we have assumed in our model), we conclude that the lower cut-off
in the observed distribution of hardnesses is a real phenomenon. That
is, there really are very few soft GRBs, and $N(H)$ does have a lower
cutoff.

The story is, however, very different for larger values of $H$.  The
data show very small numbers of hard bursts; for instance, only two
bursts out of 54 have $H> 1$MeV. Nevertheless, this does {\it not}
mean that GRBs intrinsically cut-off for hardnesses above $1$MeV.  For
instance, our theoretical model with $\xi=0$ and $\gamma=0$, with
equal number of bursts per logarithmic hardness interval, actually
predicts that even fewer bursts should have been observed. The best
fit to the data is obtained with $\gamma+3
\xi/2 =0.5$ which corresponds to a burst population with an increasing
number of bursts with hardness (for $\xi=0$) or to an equal number of
bursts per logarithmic hardness interval ($\gamma$ = 0) but with
harder bursts being more luminous ($\xi=1/3$).  


\begin{figure}
\begin{center}
\leavevmode
\epsfxsize=150pt \epsfbox{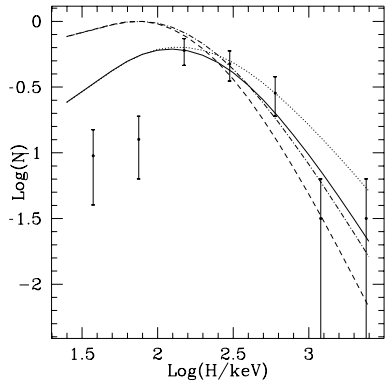}
\end{center}

\caption{Observed and expected numbers of detected bursts per interval
of $\log(H)$. The the
error bars represent statistical errors.  The best fit (solid line)
corresponds to $\alpha = \tilde \alpha= -0.41$, $\gamma+3\xi/2 =
0.5$. Other curves are: $\alpha =\bar \alpha= -0.7$, $\gamma+3\xi/2 =
0.5$ (dotted line), $\alpha = \tilde \alpha=-0.41$, $\gamma+3\xi/2 =0$ 
(dashed line), and $\alpha = \bar \alpha=-0.7$, $\gamma+3\xi/2 = 0$
(dashed-dotted line).}
\end{figure}

The interpretation of the result is quite simple. There is an
observational bias against detecting bursts with $H\ \sgreat\ 500$keV
by current detectors. Two factors operate.  For bursts with a fixed
luminosity, harder bursts have fewer photons.  This makes the
detection of harder bursts more difficult in any detector that is
triggered by photon counts. The decrease in sensitivity in BATSE is
even more severe since BATSE triggers on photons in the 50keV to
300keV range and as the bursts becomes harder most of the emitted
photons are further and further away from this energy range.

The last point depends, of course, on the power law index $\alpha$ in
the low energy range. It suggests that hard bursts that are detected
by BATSE will have values of $\alpha$ lower than average. Fig. 2
depicts the distribution of $\alpha$ values in different hardness
regimes, and shows the effect clearly.  Among the 54 bursts in the
Band sample, hard bursts do have significantly more negative values of
$\alpha$.  It is for this reason that we have introduced
$\tilde\alpha$, the average $\alpha$ value over the intermediate
hardness range, $50\,{\rm keV}<H<300\,{\rm keV}$. If the intrinsic
$\alpha$ distribution is independent of the hardness than
$\tilde\alpha$ is a better estimate of the average $\alpha$ than $\bar
\alpha$.

Figure 2, by itself, without any of the theoretical arguments
presented earlier, suggests the existence of the selection effect we
have discussed.  There are two ways of interpreting the evidence in
this plot.  One could say that the plot reflects the true distribution
of burst properties and that for some reason $\alpha$ happens to be
correlated with $H$ in the particular manner seen in the data.  This
is very ad hoc.  The alternative, which we find much more attractive,
is to say that BATSE has difficulty detecting hard bursts, and finds
it particularly difficult to detect hard bursts with less negative, or
positive, values of $\alpha$; that is, such bursts do exist, but BATSE
misses them.  If we accept the latter explanation, then it means that
BATSE is certainly missing at least some part of the hard population,
namely those bursts with less negative (or positive) $\alpha$.  It is
then but a small step to accept the entire argument presented earlier.

\begin{figure} 
\begin{center}
\leavevmode
\epsfxsize=150pt \epsfbox{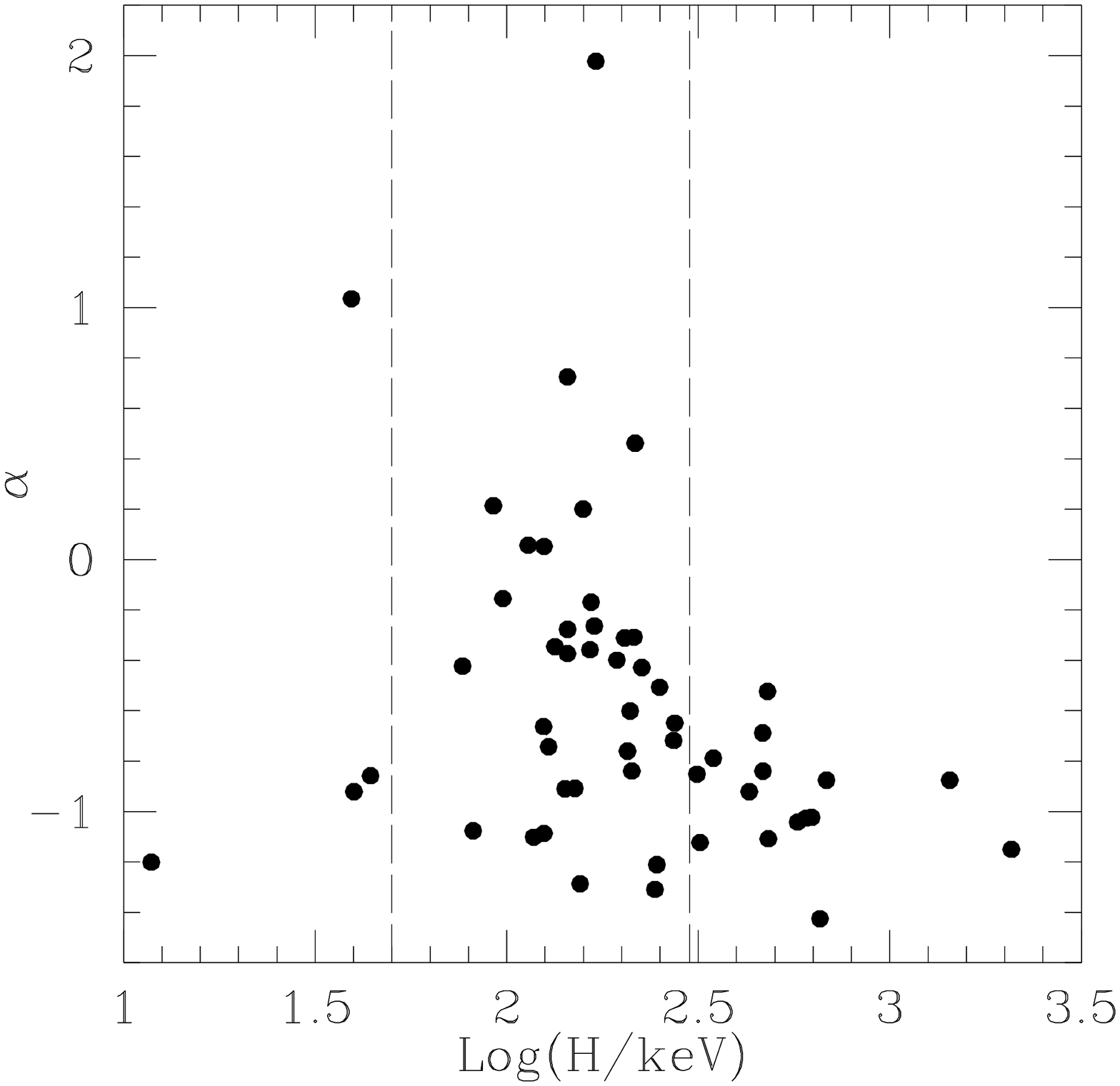}
\end{center}

\caption{The low energy spectral slope $\alpha$ vs. the hardness $H$ in
the Band sample. Hard bursts tend to have more negative
$\alpha$, which we interpret as evidence for the selection effect
discussed in the paper. The vertical lines mark 50keV and 300keV}
\end{figure}

Our main result then is the following.  When observational selection
effects are taken into account, the two hard bursts with $H>1$MeV in
the Band sample are but ``the tip of the iceberg,'' and represent a
large number of undetected hard GRBs.  The observed distribution of
hardness is, in fact, consistent with the possibility that {\it most}
GRBs are harder than the bursts detected by BATSE, and that there is a
large population of mostly undetected GRBs whose typical photon energy
is in the MeV, or even harder, range.

So far we have compared the theoretical predictions to the hardness
distribution in the Band sample. 
It will be interesting to perform the same analysis
on a larger sub-sample (see \cite{CNP}). It
will also be of great interest to perform a similar analysis on data
from other GRB detectors. One intriguing possibility is to search in
BATSE's raw channel 4 data for untriggered events, which might be some
of the missing hard bursts.  This would be a complementary 
search to
the soft GRB  search reported  by Kommers et. al. \cite{Kommers}.  In the
meantime, before we know whether hard bursts exist or not, we should
be very cautious about performing correlations between various
characteristics of the observed bursts and their hardness
parameters. The data are biased, as far as H is concerned, and any
correlation seen (e.g. the correlation between $\alpha$ and $H$ in
Fig. 2) might reflect nothing more than selection effects in the data.

We thank B. Pacz\'nski for remarks that motivated this research. The
research was supported by a NASA grant NAG5-1904 to Harvard University
by a Basic research grant to the Hebrew University
and by NSF grant PHY94-07194 to the ITP.


\begin{references}

\bibitem{Band} Band, D., et. al., 1993, Ap. J. {\bf 413}, 281.

\bibitem{DRM}  Pendleton et. al. 1995, preprint.

\bibitem{CNP} Cohen, E., Narayan, R., \& Piran, T., 1996, in preparation.

\bibitem{Kommers} Kommers,  J.M.,  Lewin, W.H.G., van Paradijs, J.
Kouveliotou C. \&  Fishman G.J, 1996, this conference.
\end{references}
\end{document}